\documentclass[eqsecnum,tightenlines,superscriptaddress,nofootinbib,10pt]{revtex4}

\usepackage[english]{babel}

\usepackage[letterpaper,top=2cm,bottom=2cm,left=3cm,right=3cm,marginparwidth=1.75cm]{geometry}

\usepackage[colorlinks=true, allcolors=blue]{hyperref}
\usepackage{algpseudocode}
\usepackage{amsmath,amsfonts,amssymb,amsthm}
\usepackage{graphicx}
\usepackage{float}
\usepackage{dsfont}
\usepackage{subcaption}
\usepackage{color}
\usepackage{multirow}  
\usepackage{empheq}
\usepackage{algorithm}
\usepackage{svg}
\usepackage{multirow}  
\usepackage[format=plain,font=small,textfont=it]{caption}
\usepackage{booktabs}
\usepackage{amsfonts}

\usepackage[colorinlistoftodos,prependcaption,textsize=footnotesize]{todonotes}

\newcommand{\aref}[1]{\hyperref[#1]{appendix~\ref*{#1}}}

\newcommand{\Wi}{W \left(   j , i_n \right)}

\newcommand{\Nf}{\frac{1}{Z}}

\newcommand{\Wthree}[6]{\left(\begin{array}{ccc} #1 & #2 & #3 \\ #4 & #5 & #6 \end{array}\right)}
\newcommand{\Wfour}[9]{\left(\begin{array}{cccc} #1 & #2 & #3 & #4 \\ #5 & #6 & #7 & #8 \end{array}\right)^{(#9)}}
\newcommand{\Wsix}[6]{\left \{ \begin{array}{ccc} #1 & #2 & #3 \\ #4 & #5 & #6 \end{array}\right \} }

\begin{document}

\title{Generative Flow Networks in Covariant Loop Quantum Gravity}

\author{Joseph Bunao}
\author{Pietropaolo Frisoni}
\affiliation{Department of Physics and Astronomy, University of Western Ontario, London, ON N6A 5B7, Canada}
\author{Athanasios Kogios}
\affiliation{Perimeter Institute, 31 Caroline Street North, Waterloo, ON, N2L 2Y5, CAN}
\affiliation{Department of Physics and  Astronomy, University of Waterloo, 200 University Avenue West, Waterloo, ON, N2L 3G1, Canada}
\author{Jared Wogan}
\affiliation{Department of Physics and Astronomy, University of Western Ontario, London, ON N6A 5B7, Canada}

\begin{abstract}
Spin foams arose as the covariant (path integral) formulation of quantum gravity depicting transition amplitudes between different quantum geometry states. As such, they provide a scheme to study the no boundary proposal, specifically the nothing to something transition and compute relevant observables using high performance computing (HPC). Following recent advances, where stochastic algorithms (Markov Chain Monte Carlo-MCMC) were used, we employ "Generative Flow Networks", a newly developed machine learning algorithm to compute the expectation value of the dihedral angle for a 4-simplex and compare the results with previous works.
\end{abstract}

\maketitle

\section{Introduction}

A universe described by a Big Bang, assumes the existence of an initial singularity (that means a manifold with a boundary), an initial state from which the expansion begins and evolves resulting (eventually) in the large scale structure we observe today. Considering this model in a quantum mechanical setting, we can view the whole process as a transition amplitude between two states: the initial singularity and the universe in a later moment. But quantum theory studies the dynamical evolution of a system given some initial conditions (defined on the boundary), which in our case, for the initial singularity, should be put by hand, as the model suggests that everything (including matter and spacetime itself) were created exactly at the moment of Big Bang.

Instead of dealing with a fine tuning issue, a rather more agnostic approach is to consider a no boundary initial state \cite{PhysRevD.28.2960} (for a pedagogic review see \cite{LEHNERS20231}) where matter and spacetime were not created (nothing). A schematic way to understand this would be to think of a $2$-dim sphere, which encompasses everything but still has no boundary. Of course, a sphere has positive curvature and would correspond to a Euclidean spacetime. Generalizing this idea to a $4$-dim spacetime, we can regard a $S^4$ as the initial state in our transition amplitude without the need to define initial conditions for our equations and inspect the transition amplitude to a state describing our current universe ie homogeneous and isotropic (something). Fortunately, there is a framework to express transition amplitudes using a path integral formulation of quantum gravity, namely spin foams.

In our case, the tunnelling process from nothing to something corresponds to what is called a $4$-simplex, with additional constraints to account for the cosmological context we are working on\footnote{A general $4$-simplex as a spin foam amplitude describes the transition between two general quantum geometries, from nothing to a 4D one}.Therefore, we require the graph to have homogeneous boundary data (homogeneity) put differently all spins in the boundary of the graph will have the same value $j$ which will be fixed for each simulation. Additionally, the graph should be regular that is to say that being in any node should not change the "view" (isotropy). Our aim will be to compute an observable on it, the dihedral angle. The latter (contrary to the usual definition) will denote the angle between two triangles in the same tetrahedron thus providing a notion of local geometry.

Despite the multiple analytical techniques developed to describe our physical environment, it is, still, a cumbersome procedure to actually perform computations and derive results. In the context of quantum gravity and, especially, spinfoams we are interested in solving complicated path integrals, which is, analytically, an almost intractable matter. As a consequence, high performance computing (HPC) techniques \cite{Francesco_draft_new_code} along with stochastic methods (mainly Markov Chain Monte Carlo/MCMC) \cite{Spinfoam_Lefschetz_thimble, Steinhaus:2024qov} were proposed as a viable alternative. In \cite{Frisoni:2022urv}, a simple spinfoam model describing the simplest triangulation of a 3 sphere emerging from a 4-simplex was studied using MCMC. Specifically, only the homogeneous sector, where all boundary links have the same spin value, was considered, therefore the only degrees of freedom (dof) come from the boundary intertwiners (nodes). Then, expectation values of some operators and their correlations were derived. This process included the numerical computation of the spinfoam amplitude in terms of the aforementioned dof and the use of a MCMC sampler. 

As explained, there is an ever increasing implementation of stochastic algorithms as a way to solve our equations in practice. Although they present a lot of advantages, MCMC methods are not free from drawbacks. To be more specific, they tend to be problematic in cases where the high probability peaks are well separated from each other, particularly in high dimensional distributions. Since the MCMC moves through the distribution in stochastic random-walk-like small perturbations finding new isolated modes becomes less probable as the algorithm focuses on the peaks it has already found. In other words, given multiple high peaks and a random initialization of the algorithm, its effectiveness on finding all peaks depends on how well the random initial points cover the distribution. Its ability to find and focus to a peak is, also, its drawback. 

To tackle this issue, some machine learning algorithms arose. To be more specific, Generative Flow Networks (GFlowNets) were developed in the last years as a possible alternative to traditional MCMC methods in order to try to gauge an unknown, complex probability distribution. \cite{Kanwar_2020, Albergo_2019, malkin2022trajectory, NEURIPS2021_e614f646, bengio2022gflownet, Bengio_tutorial, lahlou2023torchgfn}. In spite of being computationally more expensive to train than an MCMC, the cost is amortized as the network needs to be trained only once. Then, sampling a constructive sequence is faster than the corresponding MCMC one, as the latter needs to be rerun for every sampling occasion. In addition, a GFlownet is more efficient in finding the isolated peaks, as it utilizes all of the pairs of points-probabilities visited during the training procedure and, also, generalizes and infers the locations of other peaks in areas which it did not visit yet.

The paper starts with a brief introduction on the Generative Flow Networks' (GFlowNets) mathematical foundations and the different loss functions considered in our calculations, \autoref{sec:GFN_background}.  In \autoref{sec:application_spinfoam}, we shortly present the theory behind spinfoams and how to formulate them in the context of stochastic methods and as a GFlowNet hypergrid. Finally, we compare the effectiveness of the two methods and discuss the respective results.

\section{GFlowNet background}
\label{sec:GFN_background}
Generative Flow Networks (GFlowNets) are algorithms that generate composite samples according to a target unnormalized distribution via a series of assembling steps. These algorithms comprise (1) an agent that can appropriately map intermediate object states to a probability distribution of constructive actions and (2) a training objective whose global minimum enforces consistent probability flow throughout the building steps and fits the target distribution at the terminal states. Their novelty lies on the fact that instead of optimizing a problem to a single-maximizing sequence, they generate a wide variety of high-return solutions based on a reward function $R(x)$. 

\subsection{Mathematical Preliminaries}
Consider a Directed Acyclic Graph (DAG), $G=(\mathcal{S},\mathcal{A})$ where the nodes $s\in \mathcal{S}$ are all the possible states in the system being considered and the edges $a=s_i \rightarrow s_{i+1}\in \mathcal{A} \subset \mathcal{S}\times \mathcal{S}$ are the constructive actions transforming $s_i$ to $s_{i+1}$. We now define $s_0$ as the unique state without incoming edges (no parent states) and $\mathcal{S}_T \subseteq \mathcal{S}$ as the set of states without outgoing edges (no children states). Consequently, we name $s_0$ the initial state and $s_T \in \mathcal{S}_T$ a terminal state. Additionally, we call a sequence of states connecting $s_0$ and $s_T$ as a complete trajectory which we can explicitly write as $\tau = s_0\rightarrow s_1\rightarrow s_2\rightarrow ... \rightarrow s_T$. Each state transition in $\tau$ corresponds to an action in $\mathcal{A}$ with the last action transitioning a state into $s_T$ being called the terminate action.

Given a trajectory $\tau$, we can define the trajectory flow $F: \mathcal{T}$, which corresponds to the percentage of flow along this trajectory. For a state  $s$, we can define the state flow: $F(s) = \sum _{s \in \tau} F( \tau )$.

Additionally, the forward probability (remember that our graph is directed) of visiting state $s'$ starting from $s$ is: $P_F(s' | s)$, while the reverse is: $P_B (s | s')$. 

Given a target reward distribution $R: \mathcal{S}_T \rightarrow \mathbb{R}^{+}$, the goal of a GFlowNet agent is then to sample terminal states $s_T$ with probability $P(s_T) = R(s_T) / Z$ where $Z$ is some proportionality constant given as $Z = \sum _{\tau \in \mathcal{T}} F(s)$. Since a GFlowNet agent generates terminal states through the construction of complete trajectories, the terminal state probability $P(s_T)$ arises from the sum over all probabilities of trajectories ending in $s_T$. Assuming a Markovian process, the contribution of a particular trajectory $\tau$ factorizes over the forward transition probabilities $P_F(s_{i+1} | s_i)$ over all $s \in \tau$. In other words, to sample terminal states with the desired probability, the transition probabilities must satisfy
\begin{equation}
\label{eq:prop_to_r}
R(s_T) / Z = \sum_{\partial\tau = s_T} \prod_{s_i, s_{i+1} \in \tau} P_F(s_{i+1} | s_i), \forall  s_T \in \mathcal{S}_T
\end{equation}

The GFlowNet agent then estimates $P_F(s_{i+1} | s_i; \theta)$ (more precisely, its logarithm) for every edge $s_i \rightarrow s_{i+1}$ in $G$, where $\theta$ represents a set of learnable parameters. We call $P_F$ the agent’s forward policy since it governs the actions of the agent for every encountered state.

\subsection{Losses}

Neural networks are trained through optimizing an appropriately chosen loss function. For GFlowNets there is, already, a variety of them in the literature and some of them were used in this paper as different training objectives might lead to different results. 

\subsubsection{Log-Partition Variance Loss}
Introduced in \cite{zhang2023robust}, the log-partition variance loss (ZVar), initially, samples a smaller graph and a terminal state/reward function stemming from the given dataset, along with a number of trajectories ending in this terminal state, called a mini batch. On this limited situation, it then proceeds to compute the  difference between the $\log Z$ of a trajectory in this subgraph and the mean $\log Z$ from the mini batch:
\begin{equation}
    \mathcal{L}_{V} (s) = \left( \log Z(s; \theta) - \mathbb{E}_{s}[\log Z (s; \theta)] \right) ^2
\end{equation}

\subsubsection{Flow Matching Loss}

The flow matching loss (FM) \cite{NEURIPS2021_e614f646} relates the ingoing with the outgoing flow for every noninitial and nonterminal state $s$. In other words, the sum of the ingoing flows to $s$ stemming from all parent nodes $s"$ should be equal to the total outgoing flow to all children nodes $s'$ ie:

\begin{equation}
    \mathcal{L}_{FM} (s) = \left( \log \frac{\sum_{(s" \to s )\in \mathcal{A}} F_{\theta} (s", s)}{\sum_{s \to s'\in \mathcal{A}} F_{\theta} (s,s')} \right) ^2
\end{equation}

\subsubsection{Trajectory Balance}
\label{subsubsec:trajectory_balance}

The trajectory balance loss (TB) \cite{malkin2022trajectory} is concerned with the conservation of the flow along a full trajectory instead of focusing on a state. To be more exact. suppose a trajectory $\tau$: the percentage of the flow moving from the initial state $s_{0}$ to a terminal one $s_{n}$ normalized to the initial number of "particles" $Z$ should equal the corresponding backward flow along the trajectory normalized with the number of output particles ie the forward flow along the given trajectory should match the backward flow:
\begin{equation}
    \mathcal{L}_{TB} (\tau) = \left( \log \frac{Z_{\theta} \prod_{t=1}^{n} P_{F} (s_{t} | _{t-1}; \theta)}{R(x) \prod_{t = 1}^{n} P_{B} (s_{t-1}|s_{t}; \theta)} \right) ^2
\end{equation}

\subsubsection{Detailed Balance Loss}
\label{subsubsec:detailed_balance_loss}

The detailed balance loss (DB) \cite{bengio2022gflownet} ensures that for every nonterminal state $s$ the flow moving from $s$ to $s'$ should be equal to the respective backward flow ie moving from $s'$ to $s$.

\begin{equation}
    \mathcal{L}_{DB} (s,s') = \left( \log \frac{F_{\theta}(s) P_{F}(s'| s; \theta)}{F_{\theta}(s')P_{B}(s|s';\theta)} \right) ^2
\end{equation}

\subsubsection{SubTrajectory Loss}

A generalization of the detailed balance \autoref{subsubsec:detailed_balance_loss} and trajectory balance \autoref{subsubsec:trajectory_balance}, the subtrajectory balance loss (SubTB) \cite{madan2023learning} was introduced where instead of comparing the incoming and outgoing flows moving to a state, only along its immediate neighbors (parents and children), this loss function takes into consideration a larger portion of the graph.

\begin{equation}
    \mathcal{L}_{SubTB}(\tau) = \left( \log \frac{F(s_{m};\theta) \prod_{i=m}^{n-1} P_{F} (s_{i+1}|s_{i};\theta)}{F(s_{n};\theta) \prod_{i=m}^{n-1}P_{B} (s_{i}|s_{i+1};\theta)} \right) ^2
\end{equation}

As this loss function is more general than the others, one can define different weighings ie how to treat the subtrajectories. Although not all of them led to results, for the sake of completeness we will present all of them.
\begin{itemize}
    \item "DB": Same as Detailed Balance loss. The loss function is computed for every one-step transition and contributes the same irrespectively of the length of the subtrajectory.

    \item "ModifiedDB": Same as the previous but the result for every one-step transition is weighed inversely proportional to the length of the trajectory. Smaller trajectories matter more.

    \item "TB": Equivalent to Trajectory loss.

    \item "equal\_within": A main trajectory is chosen and the loss function is computed for every subtrajectory within and weighed equally. Then, each main trajectory is weighed equally.

    \item "equal": All subtrajectories of all main trajectories are weighed equally.

    \item "geometric\_within": In the main trajectory, every subtrajecotry is weighed proportionally to: $\lambda ^ {l_{sub \tau}}$, where $\lambda$ is a user decided parameter and $l_{sub \tau}$ is the length of the subtrajectory. Then, all trajectories are weighed equally.

    \item "geometric": Each subtrajectory within every main trajectory is weighed proportionally to: $\lambda ^ {l_{sub \tau}}$ along all the possible subtrajectories.
    
\end{itemize}

\section{Application to spinfoams}
\label{sec:application_spinfoam}
\subsection{Boundary state and expectation values}
\label{subsec:bound_state_exp_values}
The LQG Hilbert space associated with a graph $\Gamma$ with $L$ links and $N$ nodes is usually written as:
\begin{equation}
\label{eq:Hilbert_space_LQG}
\mathcal{H}_{\Gamma} = L_2 \left[ SU(2)^L / SU(2)^N \right] \ .
\end{equation}
From now on, we omit the $\Gamma$ subscript not to weigh down the notation. The spin network states $|j_l , i_{n} \rangle$ constitute a basis in \eqref{eq:Hilbert_space_LQG}, where $n = 1 \dots N$,  $l = 1 \dots L$\footnote{isolated indices range over multiple values}. The set $j_l$ consists of half-integer spins, while $i_n$ is an intertwiner set. Each intertwiner $i$ is a basis element of the $SU(2)$-invariant subspace of the tensor product of four representations at the corresponding node, since in 4D every vertex should be $4$-valent meaning that $4$ edges should be connected to it. In this paper, we consider homogeneous boundary data, imposing all the boundary links' spins to be equal so that $j_l = j$. Therefore, only one common spin is attached to all the links in this subspace. We write a boundary spin network state as: 
\begin{equation}
\label{eq:spin_network_symm_sector}
|j , i_n \rangle = |j, i_1 \rangle \otimes \dots \otimes |j, i_N \rangle  \ ,
\end{equation}
where $N$ is a fixed number. We consider the state $| \psi \rangle$ introduced in \cite{Gozzini_primordial}. In the Hilbert space \eqref{eq:Hilbert_space_LQG} with homogeneous boundary data, it is defined as:
\begin{equation}
\label{eq:state}
| \psi \rangle = \sum_{i_n} \Wi  | j, i_n \rangle \ ,
\end{equation}
where $\Wi$ is the LQG spinfoam amplitude in the spin network basis. In spinfoam cosmology, amplitudes with a regular boundary graph and homogeneous data can be interpreted as cosmological states \cite{Bianchi:2010zs}. The amplitude truncates the spinfoam vertex expansion from nothing into a 3-dimensional geometry, compatible with the 4-dimensional Lorentzian bulk \cite{Vidotto:2011qa}. This provides a spinfoam Lorentzian version of the Hartle-Hawking wave function of the universe \cite{Hartle:1983ai}. We refer to the spinfoam cosmology literature for the physical interpretation of such states \cite{Vidotto:2011qa, Vidotto:2010kw,Roken:2010vp,Bianchi:2011ym,Hellmann:2011jn,Kisielowski:2011vu,Livine:2011up,Kisielowski:2012yv,Rennert:2013pfa}. The same state has been considered in \cite{Markov_chain_paper} to introduce (MCMC) methods in spinfoam theories' low quantum numbers regime. In such context, it was used to investigate the cellular decomposition from one 4-simplex by splitting each boundary tetrahedra into four tetrahedra (performing a graph refinement), computing entanglement entropy, expectation values, and correlations of geometrical operators. This paper considers the boundary state \eqref{eq:state} with one of the spinfoam models studied in \cite{Markov_chain_paper}, namely the 4-simplex, described in \autoref{sec:4_simplex}.

\medskip

We focus on the dihedral angle operator. It describes the external dihedral angle between faces $a$ and $b$ on the tetrahedron dual to node $k$ giving a notion of local geometry. Its expression in the basis states \eqref{eq:state} turns out to be:
\begin{equation}
\label{eq:geom-angleformula}
\langle j, i_n  | \cos(\theta)_k | j, i_n \rangle = \frac{i_k(i_k+1) - 2j(j+1) }{2 j(j+1)} \ .
\end{equation}
The dihedral angle operator \eqref{eq:geom-angleformula} has an immediate geometrical interpretation. Furthermore, it is the simplest (diagonal) operator to compute in the spin network basis \eqref{eq:spin_network_symm_sector}. The expectation value of the same operator on the state \eqref{eq:state} has been studied in \cite{Gozzini_primordial, Markov_chain_paper}. Therefore, it is an optimal candidate for testing the Generative Flow Networks algorithms applied to spinfoam theories. The expectation value on the state \eqref{eq:state} of a local operator $O$ over node $k$ is defined as:
\begin{equation}
\label{eq:operator}
\langle O_k \rangle \equiv \frac{ \langle \psi | O_k | \psi \rangle }{\langle \psi | \psi \rangle} \ .
\end{equation}
For the dihedral angle operator \eqref{eq:geom-angleformula}, the expectation value \eqref{eq:operator} can be computed as:
\begin{equation}
\label{eq:<On>}
\langle \cos(\theta)_k \rangle = \Nf \sum_{i_n} \Wi^2 \langle j, i_n | \cos(\theta)_k | j, i_n \rangle \ ,
\end{equation}
where the normalization factor is:
\begin{equation}
\label{eq:normalization_factor}
Z \equiv \langle \psi | \psi \rangle  = \sum_{i_n} \Wi^2 \ . 
\end{equation}
\subsection{Markov Chain Monte Carlo}
\label{subsec:MCMC_spinfoams}
In this paper, we employ the Metropolis-Hastings Markov Chain Monte Carlo algorithm \cite{MH_original_paper}. A summary of the MH algorithm in its general form is reported in \aref{app:MH_algorithm}. It was originally applied to spinfoams in \cite{Markov_chain_paper} to study the low quantum numbers regime and in \cite{Spinfoam_Lefschetz_thimble} to investigate the large limit, combining it with the analytical estimates due to the stationary phase approximation. In the present context, we compare the effectiveness of MCMC used in \cite{Markov_chain_paper} to compute the expectation value of quantum observables with the Generative Flow Network algorithm described in \autoref{sec:GFN_background}. 

\medskip

To compute \eqref{eq:<On>} exploiting MCMC, we first define the normalized target distribution over the state space \eqref{eq:spin_network_symm_sector} using the spinfoam amplitude:
\begin{equation}
\label{eq:squared_prop_amp}
f_{j} \left( i_n \right) = \frac{ W^2 \left( j, i_{n} \right)}{\sum\limits_{i_n} W^2 \left( j, i_{n} \right)} \ .
\end{equation}
Each diagram has a different spinfoam amplitude, which can be numerically computed. We build a Markov chain with length $N_{mc}$ constituted by intertwiner states $[i_n]_1, [i_n]_2, \dots, [i_n]_{N_{mc}}$ using the MH algorithm. Ideally, each intertwiner state is sampled (indirectly) according to the multivariate distribution \eqref{eq:squared_prop_amp}. Each state along the chain is composed of $2j+1$ intertwiners. We sample each intertwiner from a one-dimensional truncated normal distribution rounded to integers \eqref{eq:proposal_explicit}. The center of the proposal distribution is the value of the corresponding intertwiner in the previous state, truncated between $0$ and $2j$.

\medskip

The complete algorithm implementation to store Markov chains constituted by intertwiner states, using the target distribution \eqref{eq:squared_prop_amp}, is shown in the flowchart \ref{numericalcode}. The parameters of the MCMC algorithm, the definition of the proposal distribution, and the expression of the truncated coefficients related to the proposal distributions are reported in \aref{app:MH_algorithm}.
\begin{figure}
\begin{algorithm}[H]
\caption{MCMC - Random walk}\label{numericalcode}
\begin{algorithmic}[1]
\For{$j = 1 \dots j_{max}$}
\State Choose $N_{mc}$, the batch size $B$, and the standard deviation $\sigma$ as in \autoref{tbl:MH_data}
\State Sample a random intertwiners configuration $[i_{n} ]_1$ and compute $W \left( j, [i_{n} ]_1 \right)$
\State Set initial multiplicity to $1$
\For{$s = 1 \dots N_{mc}$}
\If{s \% B = 0}
\State Store $[i_{n} ]_s$, $W \left( j, [i_{n} ]_s \right)$, and the corresponding multiplicity
%\State Set $[i_{n} ]_s \rightarrow [i_{n} ]$, $W \left( j, [i_{n} ]_s \right) \rightarrow W \left( j, [i_{n} ] \right)$  
\State Dump to disk the states and multiplicities for this batch
\State Set the multiplicity to $1$ 
\State \textbf{continue}
\EndIf
\State Generate a new state $[i_n ]$ from $[i_n ]_s$
\If{$[i_n ] = [i_n ]_s$} 
\State Increase the multiplicity by $1$
\State \textbf{continue}
\Else
\State Compute $W \left( j, [i_{n} ] \right)$ 
\State Compute $p = \textrm{min} \Bigl\{ 1 , \frac{W^2 ( j, [i_{n} ] )}{W^2 ( j, [i_{n} ]_s )} \frac{C_{0 , 2j, \sigma} ( [i_{n} ]_s)}{C_{0 , 2j, \sigma} ( [i_{n} ])} \Bigl\} $
\State Generate a uniform random number $r$ between $0$ and $1$
\If{$r < p$} 
\State Store $[i_{n} ]_s$, $W \left( j, [i_{n} ]_s \right)$, and the corresponding multiplicity
\State Set $[i_{n} ]_s \rightarrow [i_{n} ]$, $W \left( j, [i_{n} ]_s \right) \rightarrow W \left( j, [i_{n} ] \right)$  
\State Set the multiplicity to $1$
\Else
\State Increase the multiplicity by $1$
\EndIf
\EndIf
\EndFor
\EndFor
\end{algorithmic}
\end{algorithm}
\end{figure}
After running algorithm \ref{numericalcode}, we use the intertwiner draws along the chain to evaluate the expectation value \eqref{eq:<On>} by applying the Monte Carlo summation:
\begin{equation}
\label{eq:<On>_MC}
\langle \cos(\theta)_k \rangle_{mc} = \frac{1}{ N_{MC}} \sum\limits_{s=1}^{N_{mc}} \langle j , [i_k]_s | \cos(\theta)_k | j, [i_k]_s \rangle \approx \langle \cos(\theta)_k \rangle \ .
\end{equation}
With respect to \eqref{eq:<On>}, we replaced a sum over $(2j+1)^{N}$ intertwiners with a sum over $N_{mc}$ intertwiner states, where $N$ is the number of boundary nodes of the corresponding spinfoam diagram. This simplifies the calculation for many nodes, which would be impractical otherwise.

\medskip

Additionally, there is a slight difference with respect to the older implementation, as the code now is run in batches. Specifically, resembling the GFlowNet batches as explained in \autoref{sec:hypergrid}, a number of random initial points is used (batch size) and the algorithm is executed parallely for each one of them. The distribution is scanned as the step grows, mapping an ever increasing volume in the configuration space.

\section{The 4 simplex}
\label{sec:4_simplex}
In this Section, we describe the spinfoam diagrams considered in this paper: the 4-simplex. The 4-simplex (or ``5-cell") triangulation has been deeply studied and discussed in LQG literature. It has five boundary tetrahedra glued on ten triangular faces. The dual triangulation is the usual vertex spinfoam amplitude, with five boundary nodes and ten links.

\medskip

\begin{figure}[t]
    \centering
    \includegraphics[scale=0.75]{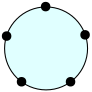}
    \caption{Schematic representation of the $4$-simplex spinfoam model considered in this paper. Each azure circle represents a vertex, each black line an edge, and each black dot a boundary node.}
    \label{fig:2-complexes}
\end{figure}

In \autoref{fig:2-complexes}, we report a schematic representation of the 2-complex spinfoam diagrams considered in this paper.

We write the EPRL vertex amplitude according to the factorization first introduced in \cite{Speziale2016}. Since then, it has been used in many papers focused on spinfoam numerical computations with the Lorentzian EPRL model \cite{Dona2019, Dona2018, Francesco_draft_new_code, Frisoni_2023, Self_energy_paper, Radiative_corrections_paper, VertexRenormalizationPaper, dona2022spinfoams, frisoni2023numerical}. We refer to \cite{Review_numerical_LQG} for an exhaustive review and description of all the steps required to switch from the holonomy representation of the EPRL vertex amplitude \cite{book:Rovelli_Vidotto_CLQG} to the one used in the numerical calculations. In the present context, we write the vertex amplitude as:
\begin{align}
\label{eq:vertex_amplitude}
W_{v}^{\gamma} \left(j , \, i_a; \, \Delta l \right) & = \sum\limits_{l_{eq} = j}^{ j + \Delta l }  \sum\limits_{k_{e}} \left( \prod_{e} (2 k_e + 1) B_{\gamma} (j,l_{eq};i_{e}, k_{e}) \right)  \{ 15 j\} \left( j, l_{eq}; k_{e}, i_{1} \right) \ ,
\end{align}
where $a = 1 \dots 5$ and $e,q = 2 \dots 5$, $e \neq q$. The integer truncation parameter $\Delta l$ cuts off the infinite sum over the auxiliary spins. These appear in the model due to the non-compactness of the $SL(2,\mathrm{C})$ group. The exact value of the amplitude is recovered only in the (ideal) limit $\Delta l \longrightarrow \infty$, although increasing such parameter considerably boosts the required computational time. The definitions of the booster function $B_{\gamma}$ and the $SU(2)$ invariant $\{ 15 j\}$ Wigner symbols are reported in \aref{app:boosters_and_Wigner}. The EPRL vertex amplitude \eqref{eq:vertex_amplitude} can be efficiently computed with the \texttt{sl2cfoam-next} numerical library \cite{Francesco_draft_new_code}. In this paper, we set the Barbero-Immirzi parameter $\gamma = 1$ and $\Delta l = 20$.

\medskip

\begin{figure}[H]
    \centering
    \includegraphics[scale=0.25]{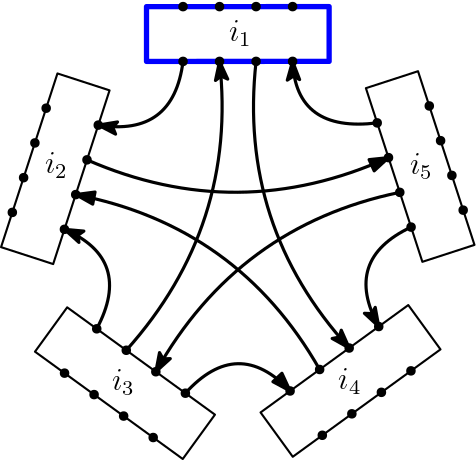}
    \caption{Wiring of the 2-complex of the vertex diagram.}
    \label{fig:2-complexes_wires}
\end{figure}

We report the vertex wiring diagram in \autoref{fig:2-complexes_wires}. It highlights the links' combinatorics. The label for each intertwiner corresponds to those used in equation \eqref{eq:vertex_amplitude}. The boundary intertwiners with a blue box are those with one $SL(2,\mathrm{C})$ integral removed to regularize the amplitude (one for each vertex). We don't explicitly label the spins not to clutter the picture.
%
\iffalse

The expression for the star spinfoam amplitude has been originally derived in \cite{Markov_chain_paper}. It can be written as the contraction of vertex tensors \eqref{eq:vertex_amplitude} as follows:
%
\begin{equation}
\label{eq:star_amplitude}
W_{\text{star}}^{\gamma} \left( j, \, i_1 \dots i_{20} ; \, \Delta l \right) = \sum\limits_{i_a}  W_{v}^{\gamma}  \left(j, \, i_a; \, \Delta l \right) \prod\limits_{b} W_{v}^{\gamma}  \left(j, \, i_b , \, i_{be}; \, \Delta l \right) \ ,
\end{equation}
%
where $b = 1 \dots 5$. To compute the corresponding amplitude, we must perform multiple contractions between vertices for each boundary data configuration. As for the vertex diagram, we report in \autoref{fig:2-complexes_wires2} the star spinfoam amplitude wiring diagram.
%
\begin{figure}[H]
    \centering
    \includegraphics[scale=0.20]{Pietro_pics/StarAmplitude.png}
    \caption{Wiring of the 2-complex of the star diagram}
    \label{fig:2-complexes_wires2}
\end{figure}
%
\fi

\section{Spinfoam Probabilities as a (GFlowNet) Hypergrid}
\label{sec:hypergrid}
The expectation value of the dihedral angle as expressed in \eqref{eq:geom-angleformula} defines a distribution over the (discrete) spin values $j$ and the intertwiners $i_{k}$. As we are working with homogeneous boundary data all links on the boundary assume the same (fixed) value of $j$, thus the only remaining variables are $i_{k}$. In this case, the number of the boundary intertwiners (which are now the only dof) determine the dimensionality of the hypergrid, which for a $4$-simplex is $5$. Each one of them can take values in the interval $[0, 2j+1]$, so we can represent the probability distribution as a discrete $5$-d dypergird, where each intertwiner is mapped to a coordinate:

\begin{equation}
    \left( i_{1}, i_{2}, i_{3}, i_{4}, i_{5}  \right),\quad i_{1}, i_{2}, \dots , i_{5} \in [0, 2j+1]
\end{equation}
and the corresponding probability amplitude is given by \eqref{eq:geom-angleformula}.

Using the GFlowNet language this discrete space defines a hypergrid, where, every point in the configuration space can be mapped to a GFlowNet state. Using this context, an agent, starting in the beginning of the grid ie for all spins equal to $0$, is set to explore it using incremental steps along each possible direction (moving only in one of them per step). Notice how we do not allow it to move in the backward direction, lowering any of the spin values. Despite not allowed to get beyond the borders of the grid (values of $j$ more than the one defined by the user), the agent can stop its scanning randomly at any point, put it in GFlowNet terms every state can be a terminal state. Subsequently, the algorithm's learning process involves getting info for the whole grid and figuring out which parts are the most relevant for our calculations. 

\medskip

\begin{figure}[h]
    \centering
    \includegraphics[scale=0.5]{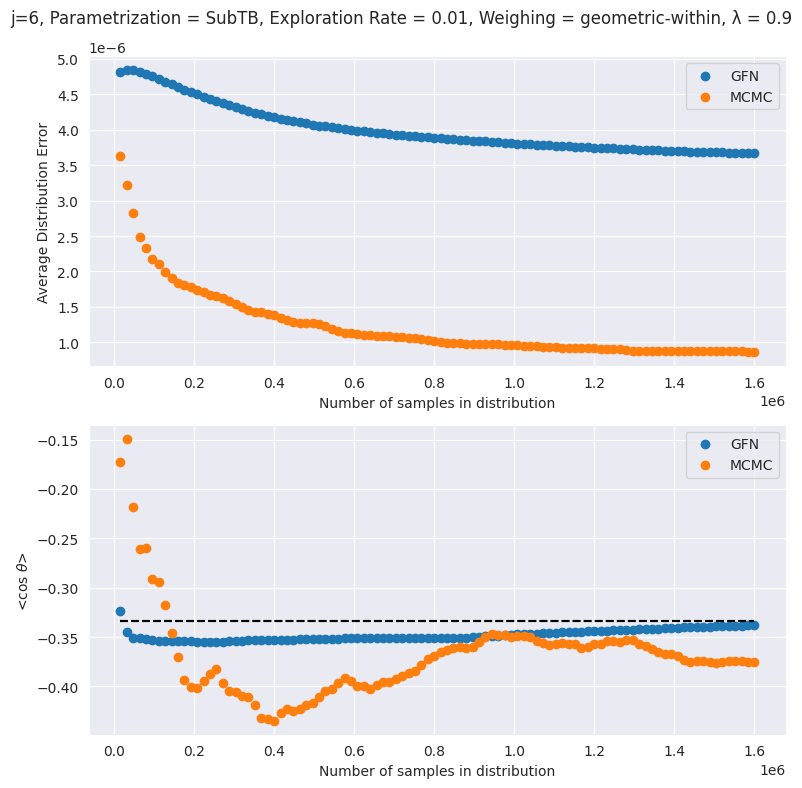}
    \caption{L1 error and observable estimation for SubTB, exploration: rate 0.01, weighing: geometric within, $\lambda$: 0.9.}
    \label{fig:Loss_Cos}
\end{figure}

\section{Results}

As discussed in \autoref{sec:GFN_background} there are multiple loss functions, we could employ in order to train our GFN. Specifically, we used: DB, TB, ZVar, FM and SubTB. For the last, we included the following weighings: DB, ModifiedDB, equal, equal within and geometric within (also, for different lambdas) . There are two additional ones, namely, TB and geometric, which did not work in our case. 

Apart from the multiple loss functions and weighings, we considered different sets of of exploration rates, thus, the number of results quickly rises. We should note that the computation time between the MCMC and the relevant GFNs differs significantly \autoref{tbl:GFN_Times}.

The GFN was composed of 2 hidden layers each coming with 256 units. It was trained in 16 batches for $1e5$ iterations each (the same numbers  was used for the corresponding MCMC calculations).

\medskip

For the analysis, we used three kinds of plots. To be more specific, on the one hand we calculated the sum of all the absolute differences in between the true value and the predicted value for both the MCMC and the GFN for the whole grid. This gives us an understanding of how much each simulation is approximating the entire grid (in a brute way). In the same figure, we included a plot presenting how close each simulation is to the observable we want to compute.

On the other, we wanted to get a more quantitative understanding of how each simulation works for each point in the grid. Since we are working in a $5D$ setting doing so is a bit tricky. Therefore, we plotted the "euclidean distance" of each point from the origin together with the predicted value.

\begin{figure}[htb]
    \centering
    \includegraphics[scale=0.4]{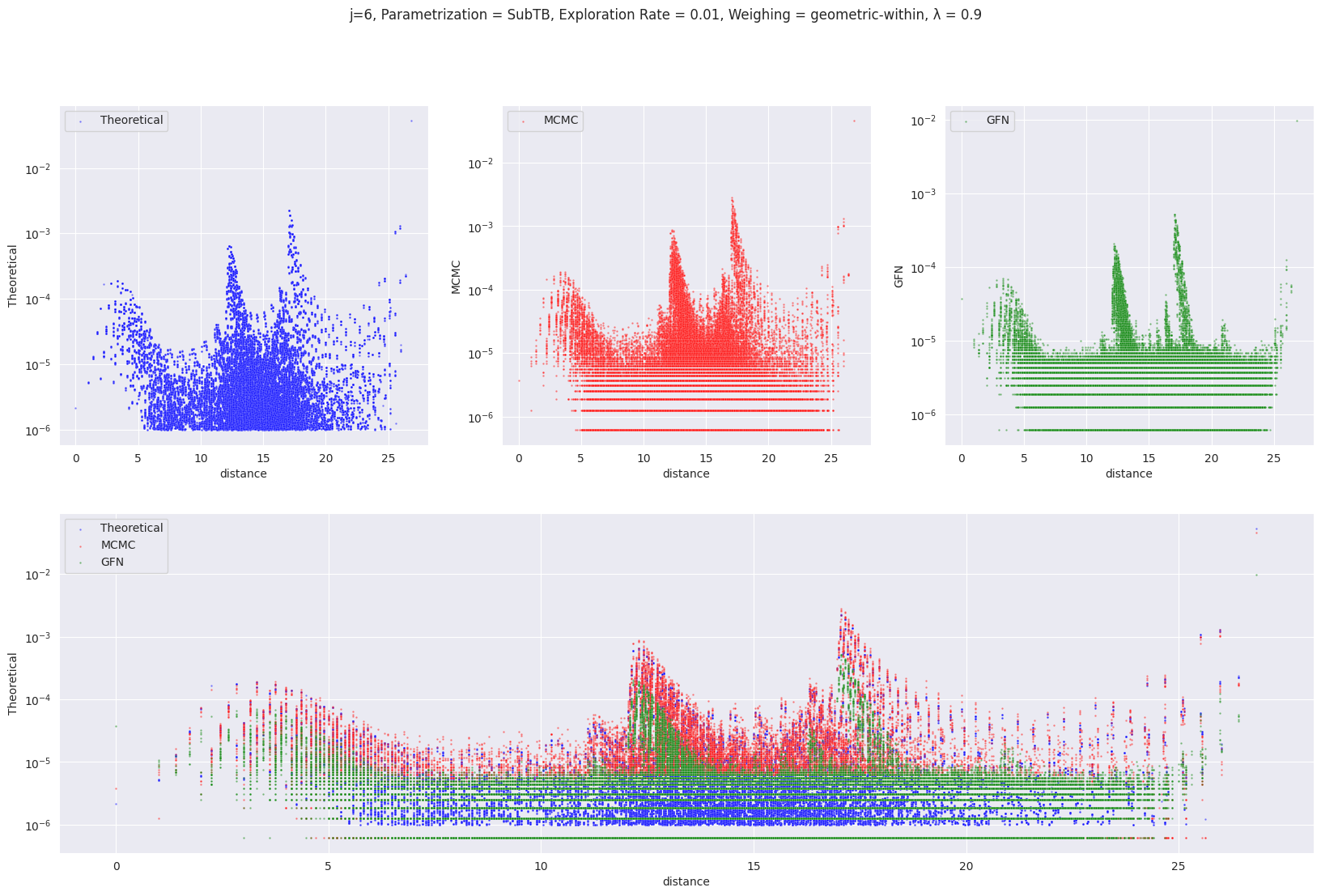}
    \caption{Euclidean distance plot for SubTB, exploration: rate 0.01, weighing: geometric within, $\lambda$: 0.9.}
    \label{fig:Euclidean_Distance}
\end{figure}

In order to keep the document short we only included one plot, which gives results comparable to the ones obtained from MCMC, to justify our later discussion. The rest can be found in the accompanying repository \cite{Repository}.

\section{Discussion}
It is evident that for some of the possible parameters' combinations, the value of the observable computed is comparable (or in some cases better) to the MCMC ones, although we should emphasize here that the corresponding training times are orders of magnitude longer \autoref{tbl:GFN_Times}. Nevertheless, both the l1 error and the "euclidean distance" one suggest that the GFN is under-performing, (despite predicting the correct location of the peaks in the latter). This contradiction can be resolved if we look closer on the philosophy of the the GFN and how they were implemented.

That is to say, GFN is used to predict the most relevant points in the grid. It moves though the latter and while doing so, learns where the most important values are and focuses on the corresponding points. As a matter of fact, that explains why in the \autoref{fig:Loss_Cos} it reaches the theoretical value faster (in terms of iterations) than the MCMC, despite the l1 error being higher. In other words, the GFN revolves around the points that contribute more to the observable we want to compute, whereas the MCMC learns more about the grid as whole (lowering the total errors).

This fact can be supported from \autoref{fig:Euclidean_Distance}. MCMC approximates better the grid in its general form, while the GFN assumes narrower peaks ie concentrates on the most dominant values.

Another argument in favor of the GFN is that it is developed as a tool to be used in cases where the peaks of the distribution are far away from each other for example higher dimensional spaces, where MCMC fails because of convergence issues (only finds some of the peaks). The problem we studied is a fairly trivial one (after all it can be solved analytically!), thus, we should expect the MCMC to perform better. Additionally as seen in \autoref{fig:Euclidean_Distance}, the peaks are not far away from each other. Nonetheless, we should focus more on the success of the algorithm finding the correct value for the observable rather than the times and resources needed, as our aim is to use it in more complicated cases where MCMC fails or would need significantly more (possibly infinite) resources. 

Furthermore, a limitation of our implementation lies on the fact that while the MCMC starts from a different point in the grid each time. The GFN always starts from 0 only moving forward, requiring longer routes (more resources) to investigate the grid. There are, though, multiple environments defined in the literature such as the EBM \cite{zhang2022generative} in which only the states in the border can be terminal state, however the agent can move freely in the grid. A future direction would be to follow and, possibly, mix these different GFlowNet constructions so that the algorithm decides on itself its next steps (without any restrictions on the orientation) freely finding possibly better (and, also, faster) the high peak areas. In addition, the exploration could start from the middle (shorter routes) or a random point every time. 

Additionally, another interesting direction would be to use a hybrid algorithm. MCMC is way faster when the peak is located, whereas GFN is better at finding where the peaks are in complex situations. Combining both of them could help make the computations faster and more accurate.

Finally, concerning resources, we have used a limited number of batches as well as hidden dimensions and layers for our simulations restricting the learning capabilities of our neural network. Therefore, it would be interesting to check how the results depend on these parameters.

The previous ideas were all based on using the GFlowNets in the problem in discussion ie calculating expectation values for operators in the context of spinfoam diagrams. Another possibility would be to apply the algorithm to building new graphs. As mentioned earlier, spinfoams are essentially graphs describing quantum geometry in a path integral format. It is, therefore useful to know which states in the phase space contribute more and a way to access these in a stochastic way. Since GFlowNets were initially used for the formation of drug molecules \cite{NEURIPS2021_e614f646} an analogous constructive procedure could be implemented, so that the agent predicts which state will be more probable during the sampling process.

\begin{acknowledgments}
 The authors acknowledge the Anishinaabek, Haudenosaunee, L\=unaap\'eewak, Attawandaron, and neutral peoples, on whose traditional lands Western University and the Perimeter Institute are located.

The research of PF and JW at Western University is supported by the by the Natural Science and Engineering Council of Canada (NSERC) through the Francesca Vidotto’s Discovery Grant ``Loop Quantum Gravity: from Computation to Phenomenology”  and by the ID\# 62312 grant from the John Templeton Foundation, as part of the project \href{https://www.templeton.org/grant/the-quantum-information-structure-ofspacetime-qiss-second-phase}{``The Quantum Information Structure of Spacetime'' (QISS)}. 

AK is supported by an NSERC grant awarded to Lee Smolin. Research at Perimeter Institute is supported in part by the Government of Canada through the Department of Innovation, Science and Economic Development Canada and by the Province of Ontario through the Ministry of Colleges and Universities.

\end{acknowledgments}

\clearpage

\appendix 

\section{Metropolis-Hastings algorithm}
\label{app:MH_algorithm}
The Metropolis-Hastings algorithm \cite{MH_original_paper} is a Markov Chain Monte Carlo algorithm that generates samples from a high-dimensional target probability distribution. We briefly review the algorithm by considering a multidimensional discrete variable $x$ on a state space $\chi$. We define $W^2_{\chi}(x)$ as an unnormalized target distribution on $\chi$. We refer to the normalized target distribution as follows:
\begin{equation}
\label{eq_app:normalized_target}
f_{\chi}(x) \equiv \frac{W^2_{\chi} \left( x \right)}{\sum\limits_{x \in \chi} W^2_{\chi}\left( x \right)} \ .
\end{equation} 
If we interpret $\chi$ as a (discrete) Hilbert space, we can compute the expectation value of a quantum observable $O$ on $\chi$ using the normalized target distribution \eqref{eq_app:normalized_target} as follows:
\begin{equation}
\label{eq_app:operator_to_compute}
O = \frac{ \sum\limits_{ x \in \chi} W^2_{\chi} \left( x \right) o \left( x \right) }{\sum\limits_{x\in \chi} W^2_{\chi}\left( x \right)} \ ,
\end{equation} 
where $o \left( x \right)$ represents the observable's matrix element in a quantum basis of the Hilbert space $\chi$. The Metropolis-Hastings algorithm allows constructing an ergodic Markov chain on $\chi$ with finite length $N_{mc}$ such that each state is (indirectly) sampled from the normalized target distribution \eqref{eq_app:normalized_target}. The condition for this is that $W_{\chi}(x)$ can be computed up to a multiplying constant. A positive proposal distribution $g_{\chi}$ on the space $\chi$ is required to transit from each state to the next one. We denote such Markov chain as: $[x]_{1}, [x]_{2}, \ \dots, [x]_{N_{mc}}$ where $[x]_s$ indicates the $s$-th state along the chain.

\medskip

In this paper, we employ the random walk Metropolis-Hastings. The sampler locally explores the neighborhood of the current state $[x]_s$ of the Markov chain, proposing a candidate state $[x]'$ sampling from the proposal distribution $g_{\chi}$. Therefore, the algorithm suggests a candidate state depending on the chain's current state $[x]_s$. The distribution $g_{\chi}$ is multivariate if the space $\chi$ is multi-dimensional (each component of the state $x$ is sampled from a one-dimensional distribution independent of the others). We consider a multivariate truncated normal distribution rounded to integers as a proposal distribution in the Metropolis-Hastings algorithm. Each component of $x$ is sampled from a one-dimensional truncated normal distribution rounded to integers:
\begin{equation}
\label{eq:proposal_explicit}
g_{n_1 , n_2, \sigma} \left( n \right) = \frac{\Phi_{\sigma} (n + 0.5) - \Phi_{\sigma} (n - 0.5)}{\sum\limits_{k = n_1}^{n_2} \left[ \Phi_{\sigma} (k + 0.5) - \Phi_{\sigma} (k - 0.5) \right]} \ ,
\end{equation}
where $\Phi_{\sigma} (x)$ is the cumulative distribution function of a one-dimensional normal distribution with zero mean and standard deviation $\sigma$. Its expression is:  
\begin{equation}
\Phi_{\sigma} (x) = \frac{1}{\sigma \sqrt{2 \pi}} \int_{- \infty}^{x} e^{-\frac{t^2}{2 \sigma^2}} dt \ .
\end{equation}
For convenience, we define the truncated coefficients:
\begin{equation}
\label{eq:truncated_coeff}
C_{n_1 , n_2, \sigma} \left( [x] \right) = \prod\limits_{i=1}^{N} \left \{  \sum\limits_{n=n_1-x_i}^{n_2-x_i} \left[ \Phi_{\sigma} (n + 0.5) - \Phi_{\sigma} (n - 0.5) \right] \right \}\ ,
\end{equation}
where we defined with $x_i$ the $i$-th component of the draw $[x]$. The Markov Chain is built by evaluating the ratio between the target distribution \eqref{eq_app:normalized_target} times the truncated coefficients \eqref{eq:truncated_coeff} at the proposal state and the same quantity at the current state. Then, we accept the proposal state with a probability equal to this state. Otherwise, we stay at the current point. The initial steps of the algorithm are usually removed as burn-in iterations during the thermalization phase, although in this paper, we set the burn-in parameter to zero.

\medskip

After storing a Markov chain with the desired length $N_{mc}$, we can compute \eqref{eq_app:operator_to_compute} using a standard Monte Carlo approach on the multi-dimensional sum over $x$. We consider all the states of the chain as a statistical sample in the Monte Carlo evaluation. We obtain a reasonably good estimate of the original quantity if the number of samples is large enough: 
\begin{equation}
\label{eq:app_O_NMC}
O_{{mc}} = \frac{1}{N_{mc}} \sum\limits_{s = 1}^{N_{mc}} o \left([x]_s \right) \approx O \hspace{3mm} \textrm{for $N_{mc} \gg 1$} \ .
\end{equation}
The Monte Carlo estimate \eqref{eq:app_O_NMC} is precisely equal to \eqref{eq_app:operator_to_compute} only in the (ideal) limit of an infinite number of samples. Moreover, the convergence is faster than the standard version of Monte Carlo \cite{VertexRenormalizationPaper} (in which the draws are sampled randomly) because each draw is generated from the distribution \eqref{eq_app:normalized_target}. 

\medskip

There is a non-zero correlation between $[x]_s$ and $[x]_{s+d}$ where $d \geq 1$. This is because each proposed state depends on the previous one (as the process is Markovian). For each quantity \eqref{eq:app_O_NMC} we can compute the autocorrelation function with lag $d$: 
\begin{equation}
\label{eq:autocorrelation}
R_{O} \left( d \right) = \frac{\sum\limits_{s=d+1}^{N_{mc}} \left( o([x]_s) - O_{N_{mc}} \right) \left(  o([x]_{s-d}) - O_{N_{mc}} \right)}{\sum\limits_{s=1}^{N_{mc}} \left( o([x]_s) - O_{N_{mc}} \right)^2} \ .
\end{equation}
Since the Markov Chain converges to a stationary distribution, the autocorrelation \eqref{eq:autocorrelation} should decrease as the lag $d$ increases.

\medskip

Finally, we report in \autoref{tbl:MH_data} the parameters used in the MCMC algorithm \ref{numericalcode} for the calculations considered in this paper.

\begin{table}[H]
\begin{center}
\begin{tabular}{|p{1cm}|p{1cm}|p{1cm}|p{1cm}|p{1cm}||}
 \hline
 \multicolumn{4}{|c|}{MH - parameters (4-simplex)} \\
 \hline
 $j$ & $N_{mc}$ & B & $\sigma$ \\
 \hline
 0.5   & $10^7$ & $10^6$ & $0.8$ \\
 1.0   & $10^7$ & $10^6$ & $0.8$ \\
 1.5   & $10^7$ & $10^6$ & $0.8$ \\
 2.0   & $10^7$ & $10^6$ & $0.8$ \\
 2.5   & $10^7$ & $10^6$ & $0.8$ \\
 3.0   & $10^7$ & $10^6$ & $0.8$ \\ 
 3.5   & $10^7$ & $10^6$ & $0.8$ \\ 
 4.0   & $10^7$ & $10^6$ & $0.8$ \\ 
 4.5   & $10^7$ & $10^6$ & $0.8$ \\
 5.0   & $10^7$ & $10^6$ & $0.8$ \\ 
 5.5   & $10^7$ & $10^6$ & $0.8$ \\ 
 6.0   & $10^7$ & $10^6$ & $0.8$ \\  
 \hline
\end{tabular}
\quad
\hskip12mm 
\begin{tabular}{|p{1cm}|p{1cm}|p{1cm}|p{1cm}|p{1cm}||}
 \hline
 \multicolumn{4}{|c|}{MH - parameters (star)} \\
 \hline
 $j$ & $N_{mc}$ & B & $\sigma$ \\
 \hline
 0.5   & $10^7$ & $10^6$ & $0.8$ \\
 1.0   & $10^7$ & $10^6$ & $0.8$ \\
 1.5   & $10^7$ & $10^6$ & $0.8$ \\
 2.0   & $10^7$ & $10^6$ & $0.8$ \\
 2.5   & $10^7$ & $10^6$ & $0.8$ \\
 3.0   & $10^7$ & $10^6$ & $0.8$ \\ 
 3.5   & $10^7$ & $10^6$ & $0.8$ \\ 
 4.0   & $10^7$ & $10^6$ & $0.8$ \\ 
 4.5   & $10^7$ & $10^6$ & $0.8$ \\
 5.0   & $10^7$ & $10^6$ & $0.8$ \\ 
 5.5   & $10^7$ & $10^6$ & $0.8$ \\ 
 6.0   & $10^7$ & $10^6$ & $0.8$ \\  
 \hline
\end{tabular}
\end{center}
\caption{\textit{Parameters used in the MCMC algorithm \ref{numericalcode}.} From left to right: \textit{$j$ is the boundary spin of spinfoam amplitude, $N_{mc}$ is the total number of iterations over the chain, ``B" is the number of states in each batch, and $\sigma$ is the standard deviation of the proposal distribution.}}
 \label{tbl:MH_data}
\end{table}
\section{Wigner symbols and booster function}
\label{app:boosters_and_Wigner}
This Appendix reports the expression of some invariant Wigner symbols and the booster function. These appear in the vertex amplitude expression \eqref{eq:vertex_amplitude}.
\subsection{Wigner symbols}
First, we write the expression of the $(4jm)$ Wigner symbols. These can be defined as the contraction of two $(3jm)$ Wigner symbols labeled by one additional spin:
\begin{equation}
\label{eq:4jm_symbol}
\Wfour{j_1}{ j_2}{ j_3} {j_4}{m_1}{m_2}{m_3}{m_4}{k}
 = \sum_{m} (-1)^{k-m} \Wthree{ j_1}{ j_2}{ k} {m_1}{m_2}{m} \Wthree{k}{ j_3}{ j_4} {-m}{m_3}{m_4} \ .
\end{equation}
Second, we report the canonical definition of the $\{6j\}$ symbol in terms of $(3jm)$ symbols, which is:
\begin{align}
\label{eq:6j_symbol}
  \Wsix{j_1}{j_2}{j_3}{j_4}{j_5}{j_6} 
  & = \sum\limits_{m_1 \dots m_6} (-1)^{\sum\limits_{k=1}^{6} ( j_k - m_k ) } \Wthree{j_1}{j_2}{j_3}{m_1}{m_2}{-m_3} \Wthree{j_1}{j_5}{j_6}{-m_1}{m_5}{m_6} \times \nonumber \\
  & \hspace{6mm} \times \Wthree{j_4}{j_5}{j_3}{m_4}{-m_5}{m_3} \Wthree{j_4}{j_2}{j_6}{-m_4}{-m_2}{-m_6} \ .
\end{align}
The $\{6j\}$ Wigner symbol \eqref{eq:6j_symbol} can be computed and stored efficiently using high-performance computing libraries such as \cite{johanssonFastAccurateEvaluation2015}. Finally, write the $\{15j\}$ symbol of the first kind according to the conventions of \cite{GraphMethods}. It can be written using \eqref{eq:6j_symbol} as follows:
\begin{equation}
    \label{eq:15jsymbol}
    \begin{split}
        \left \{ \begin{array}{ccccc} 
        j_1 & j_2 & j_3 & j_4 & j_5  \\  
        l_1 & l_2 & l_3 & l_4 & l_5  \\ 
        p_1 & p_2 & p_3 & p_4 & p_5 
        \end{array}\right \} & = 
        (-1)^{\sum_{k=1}^5 j_k + l_k + p_k} \sum_x (2 x +1) 
        \Wsix{j_1}{p_1}{x}{p_2}{j_2}{l_1} \Wsix{j_2}{p_2}{x}{p_3}{j_3}{l_2}\\ 
  & \hspace{6mm} \times \Wsix{j_3}{p_3}{x}{p_4}{j_4}{l_3} \Wsix{j_4}{p_4}{x}{p_5}{j_5}{l_4} \Wsix{j_5}{p_5}{x}{j_1}{p_1}{l_5}  \ .
    \end{split}
\end{equation} 
We refer to \cite{GraphMethods, book:varshalovic} for the analytical definitions and graphical representations of the generalized $\{nj\}$ invariant symbol appearing in the $SU(2)$ recoupling theory.
\subsection{Booster function}
The booster function encodes the contribution due to the boosts in the Lorentzian EPRL model. It was first introduced in \cite{Speziale2016} and is defined as an integral over the boost rapidity parameter $r$ in the $\gamma$-simple $SL(2,\mathrm{C})$ unitary representation:
\begin{equation}
\begin{split}
  \label{eq:boosterdef}
  B_{\gamma} &\left( j_1,j_2,j_3,j_4, l_1,l_2,l_3,l_4 ; i,k\right) 
  =\\
  &\sum_{ m_f } 
  \left(\begin{array}{cccc} l_1 & l_2 & l_3 & l_4 \\ m_1 & m_2 & m_3 & m_4 \end{array}\right)^{(k)}
  \left(\int_0^\infty d r \frac{1}{4\pi}\sinh^2r \, \bigotimes_{f=1}^4 d^{\gamma j_f,j_f}_{l_f j_f m_f}(r) \right)
  \left(\begin{array}{cccc} j_1 & j_2 & j_3 & j_4 \\ m_1 & m_2 & m_3 & m_4 \end{array}\right)^{(i)} \ ,
\end{split}
\end{equation}
where the $(4jm)$ Wigner symbol has been defined in \eqref{eq:4jm_symbol}. The expression of the reduced matrix elements $d_{jlm}^{\gamma j , j}(r)$ has been originally written in \cite{Ruhl:1970fk}, and it reads:
\begin{equation}\label{eq:dSL2C}
\begin{split}
    d^{(\gamma j,j)}_{jlp}(r) =&  
    (-1)^{\frac{j-l}{2}} \frac{\Gamma\left( j + i \gamma j +1\right)}{\left|\Gamma\left(  j + i \gamma j +1\right)\right|} \frac{\Gamma\left( l - i \gamma j +1\right)}{\left|\Gamma\left(  l - i \gamma j +1\right)\right|} \frac{\sqrt{2j+1}\sqrt{2l+1}}{(j+l+1)!}  \\
    & \times \left[(2j)!(l+j)!(l-j)!\frac{(l+p)!(l-p)!}{(j+p)!(j-p)!}\right]^{1/2} e^{-(j-i\gamma j +p+1)r} \\
    & \times \sum_{s} \frac{(-1)^{s} \, e^{- 2 s r} }{s!(l-j-s)!} \, {}_2F_1[l+1-i\gamma j,j+p+1+s,j+l+2,1-e^{-2r}] \ ,
\end{split}
\end{equation}
where ${}_{2}F_{1}$ is the Gauss hypergeometric function.

\section{GFN parameters and Times}

In the following table, we present the times needed for the training of the algorithm for the different sets of the parameters. The simulations were run in 40 cores (each core supports 2 "hyper-threads") and 200 GB of memory in Perimeter Institute's computer cluster. For comparison, in the same nodes the MCMC needed (for the same parameters) is $33.67 s$.

\begin{table}[h]
    \centering
    \begin{tabular}{cccccc}
    \toprule
    Parametrization & Exploration Rate & Weighing & $lambda$ & Time (in s) \\
    \midrule
    DB & 0.01 &   &   & 24233.27 \\
DB & 0.10 &   &   & 25473.84 \\
DB & 0.90 &   &   & 19307.97 \\
FM & 0.01 &   &   & 28059.27 \\
FM & 0.10 &   &   & 28797.55 \\
FM & 0.90 &   &   & 18619.80 \\
ZVar & 0.01 &   &   & 22933.50 \\
ZVar & 0.10 &   &   & 23698.39 \\
ZVar & 0.90 &   &   & 17302.39 \\
SubTB & 0.01 &  DB &   & 27629.19 \\
SubTB & 0.01 &  DB &   & 34006.91 \\
SubTB & 0.10 &  DB &   & 28957.59 \\
SubTB & 0.10 &  DB &   & 35847.32 \\
SubTB & 0.90 &  DB &   & 21059.68 \\
SubTB & 0.01 &  ModifiedDB &   & 28199.58 \\
SubTB & 0.01 &  ModifiedDB &   & 28557.65 \\
SubTB & 0.10 &  ModifiedDB &   & 30001.29 \\
SubTB & 0.10 &  ModifiedDB &   & 30329.59 \\
SubTB & 0.90 &  ModifiedDB &   & 20081.74 \\
SubTB & 0.90 &  ModifiedDB &   & 20226.71 \\
SubTB & 0.01 &  equal &   & 26841.31 \\
SubTB & 0.01 &  equal &   & 26309.74 \\
SubTB & 0.10 &  equal &   & 28822.90 \\
SubTB & 0.10 &  equal &   & 27456.59 \\
SubTB & 0.90 &  equal &   & 21543.87 \\
SubTB & 0.90 &  equal &   & 21971.17 \\
SubTB & 0.01 &  equal\_within &   & 25821.10 \\
SubTB & 0.10 &  equal\_within &   & 26651.84 \\
SubTB & 0.90 &  equal\_within &   & 20849.78 \\
SubTB & 0.01 &  geometric\_within &  0.1 & 27996.97 \\
SubTB & 0.01 &  geometric\_within &  0.9 & 25057.91 \\
SubTB & 0.01 &  geometric\_within &  3 & 25803.05 \\
SubTB & 0.10 &  geometric\_within &  0.1 & 29644.58 \\
SubTB & 0.10 &  geometric\_within &  0.9 & 26601.28 \\
SubTB & 0.10 &  geometric\_within &  3 & 26717.54 \\
SubTB & 0.90 &  geometric\_within &  0.1 & 20455.08 \\
SubTB & 0.90 &  geometric\_within &  0.9 & 20818.05 \\
SubTB & 0.90 &  geometric\_within &  3 & 21188.65 \\
    \bottomrule
    \end{tabular}
    \caption{Times for different parameters.}
    \label{tbl:GFN_Times}
\end{table}

\clearpage

\bibliographystyle{utcaps}
\bibliography{PP_references, sample}

\end{document}